# STATUS OF POLARIZATION CONTROL EXPERIMENT AT SHANGHAI DEEP ULTRAVIOLET FREE ELECTRON LASER*

Haixiao Deng[#], Tong Zhang, Lie Feng, Bo Liu, Jianhui Chen, Zhimin Dai, Yong Fan, Chao Feng
Yongzhou He, Taihe Lan, Dong Wang, Xingtao Wang, Zhishan Wang, Jidong Zhang
Meng Zhang, Miao Zhang, Zhentang Zhao, SINAP, Shanghai 201204, China
Lin Song, BUAA, Beijing100191, China

*Abstract*

A polarization control experiment by utilizing a pair of crossed undulators has been proposed for the Shanghai deep ultraviolet free electron laser test facility. Numerical simulations indicate that, with the electromagnetic phase-shifter located between the two crossed planar undulators, fully coherent radiation with 100 nJ order pulse energy, 5 picoseconds pulse length and circular polarization degree above 90% could be generated. The physical design study and the preparation status of the experiment are presented in the paper.

## INTRODUCTION

With the rapid development of accelerators and other related techniques, scientists now have the ability to investigate into realms with much smaller and faster scale, especially thanks to the great success of x-ray FEL light sources, LCLS [1] and SACLA [2]. On the other hand, in order to study the regime such as electron spins, bonding dynamics and the valence charge [3] etc., the light sources with controllable polarization is necessary. Taking the great advantages of FEL, scientists can obtain the light sources with both powerful and polarization controllable coherent radiations within the full spectral regime from far infrared to the hard x-ray. Conventionally, polarization controllable light can be directly made from the helical undulators, e.g. APPLE-type [4], at the 3$^{rd}$ synchrotron radiation light sources [5]. However, it is quite difficult to achieve fast helicity switching by the abovementioned approach.

One possible solution for fast helicity switching is the technique of crossed planar undulators [6]. The so called crossed planar undulators are a pair of planar undulators with the orthogonal magnetic orientation, thus providing a combined radiation with various polarizationu. The crossed undulators have been successfully utilized to generate the circularly polarized radiation at storage ring-based FELs [7, 8], and was proposed at high-gain FELs [9-14].

In the following sections, we report the recent progress of the polarization control FEL experiments proposed at Shanghai deep ultraviolet free electron laser (SDUV-FEL) test facility, both the physical designs and the hardware status will be covered.

## PHYSICAL DESIGN AND SIMULATION

The SDUV-FEL is a versatile test bed, well suited for testing novel FEL principles [15]. Up to now, various great successes have been achieved on this machine experimentally, including SASE [16], HGHG saturation [17] and first lasing of EEHG [18]. In order to carrying out the FEL polarization control experiment, actually a proof-of-principle experiment for the future x-ray FEL, a few hardware modifications have to be performed.

The schematic layout of the FEL polarization control experiment at SDUV-FEL is shown in Figure 1 [19], from which, a pair of planar undulators with crossed magnetic orientation and phase-shifter between them can be found.

The three-dimensional start-to-end simulation has been accomplished with the codes ASTRA [20], ELEGANT [21] and GENESIS [22]. The electron beam with energy of ~136MeV interacts with the seed laser (1047 nm) in the modulator (EMU65), so as to form beam energy modulation. The dispersive Chicane-I is introduced to convert the energy modulation into longitudinal density modulation, so as to enhance the harmonic bunching (here is 523 nm, i.e. 2$^{nd}$ harmonic of the seed laser).

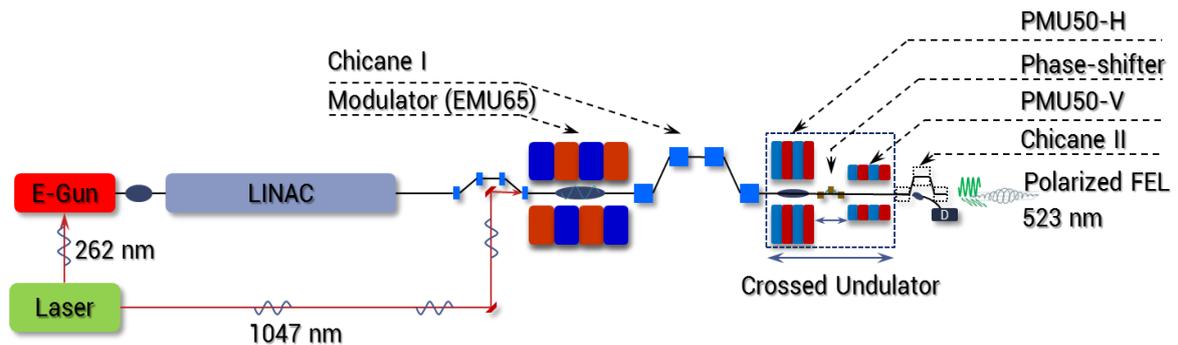

Figure 1: The schematic layout of polarization control FEL experiments at SDUV-FEL.

___________________
*Supported by NSFC (11175240) and 973 Program (2011CB808300).
#denghaixiao@sinap.ac.cn

With a pair of crossed undulators downstream, coherent FEL radiations with orthogonal polarized orientation could be achieved. Additionally, tuning the strength of the phase-shift between the two crossed planar undulators, the circular polarization control of the final combined radiation could be also realizable. In principle, as the tuning of the electromagnetic phase-shifter is so easy, that the fast polarity switching could be fulfilled.

The preliminary physical design and calculations can be reached from ref. [19]. Here the upgraded simulation results are presented according to the recently frozen experiment configuration. In Figure 2, it is apparently that with tuning of the phase-shifter, circular polarized FEL could be varied between the maximum (~90%) and the minimum values accordingly. Furthermore, the final output FEL radiations show good coherence and narrow bandwidth which inherent from the seed laser.

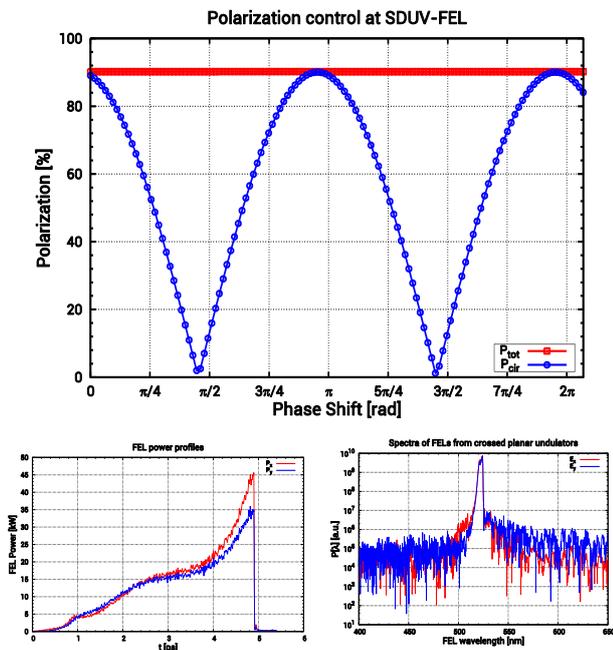

Figure 2: Start-to-end simulation of the FEL polarization control with the frozen configuration at SDUV-FEL (upper panel). The output FEL radiations with both x and y directions in time and frequency domain, respectively (lower panel).

## STATUS

As it is mentioned before, only some minor hardware modifications needs to be performed, in order to make the SDUV-FEL suit for FEL polarization control experiment. The configuration under upgrading is listed as following:
- A new customized permanent planar undulator (PMU50-V) whose magnetic field orientation is orthogonal with the existing one (PMU50-H);
- A small phase-shifter section which consists of three electromagnetic dipoles;
- The layout configure updating at SDUV-FEL, i.e. rearrangement of the devices on the SDUV-FEL beam line to form the optimal experiment setup according to the simulation results;
- The preparation of the optical diagnostics site for the measurements of the polarization performance of the FEL lights;
- Other related issue, e.g. the control panel of newly added hardware sections, etc.

Up to now, the PMU50-V which is responsible for the vertical polarized FEL is on the components assembly stage. The undulator girder and the transmission system are ready, seen in Figure 3. Once the magnetic blocks are measured, the whole undulator manufacture will be completed around the beginning of this September, and the next plan will be magnetic field measurements and calibrations.

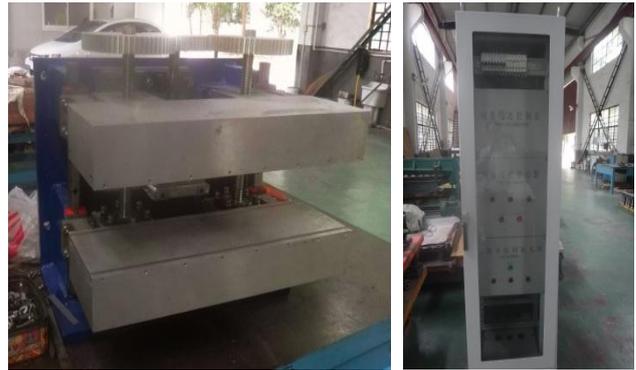

Figure 3: PMU50-V girder (left) and transmission control system (right).

The phase-shifter manufacture was completed, seen in Figure 4, and the magnetic field is under measurement.

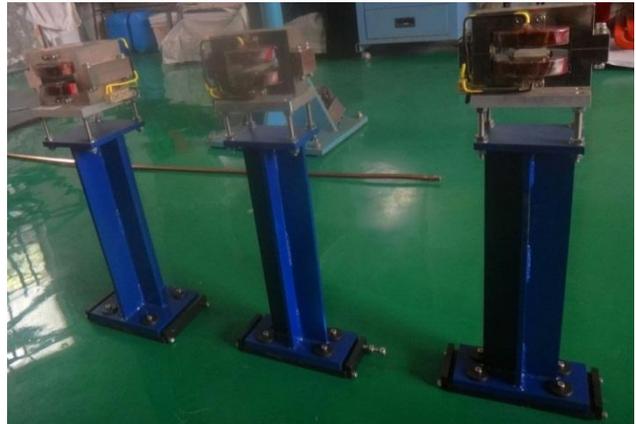

Figure 4: Phase-shifter section between the two crossed planar undulators, which is made up of 3 electromagnetic dipoles.

Other components, e.g. the modified vacuum chambers, bellow pipes and various holders will be ready by the end of this August. Once all the elements are ready, final installation on SDUV-FEL will be carried out around this October, i.e. two months later. Figure 5 illustrates the hardware updating from the present configuration to the optimal one for the polarization control FEL experiments.

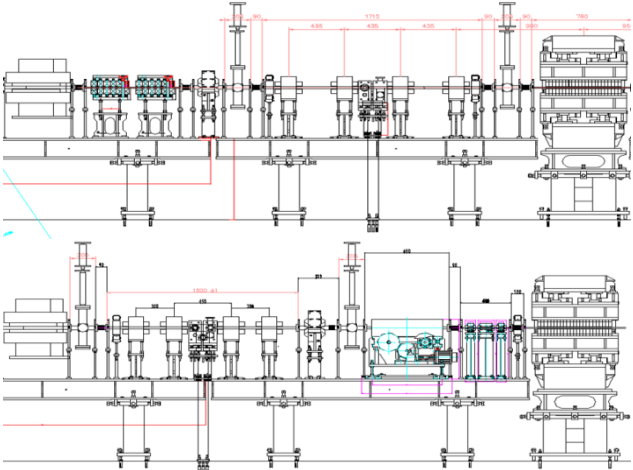

Figure 5: The CAD designs of the part of SDUV-FEL for polarization FEL control experiments, the configuration at present (upper) and the modified one for the experiments (lower).

## OPTICS DIAGNOSTICS

One of the most important issues of this experiment is the measurement of polarized status of the final output FEL radiations. In ref. [19], we have already proposed a method for polarization measurements. The proposal is based on the basic optic principle, i.e. to figure out the 4 Stokes parameters [23], thus calculate the polarization degrees. The optical diagnostics station has been already finished design and installation, now the workbench is in the drive-laser hutch for off-site testing and calibration (see Figure 6).

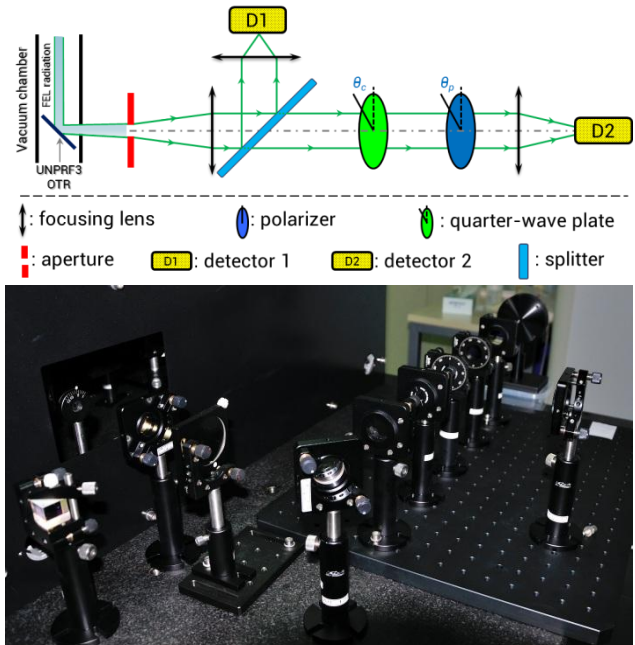

Figure 6: Optical diagnostics station, schematic layout (upper) and photograph (lower)

The off-line calibrations of the optical workbench have already been finished by using a standard 523 nm laser, i.e., the 2nd harmonic of the photocathode driven laser. By tuning the polarizer and the quarter-wave plate, several groups of intensity can be obtained by energy-meter, from which Stokes parameters can be calculated, then the total polarization and circular polarization degrees. Figure 7 shows us two groups of the measurement results, and the corresponding calibrated results can be found in Table 1.

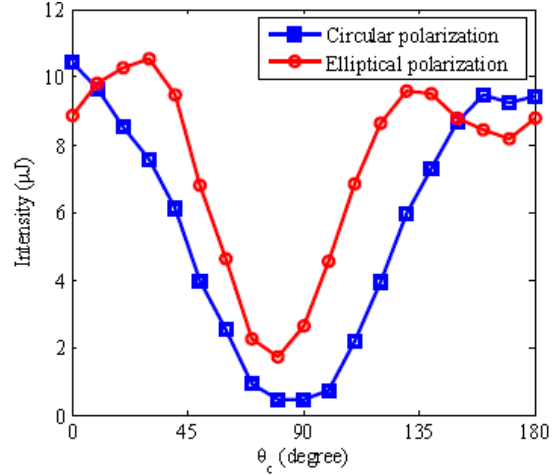

Figure 7: Two groups of off-line calibration results at the optical diagnostics station.

Table 1: The optical diagnostics off-line calibrations.

|  | $S_0$ | $S_1$ | $S_2$ | $S_3$ | $P_{circular}$ |
|---|---|---|---|---|---|
| Circular | 1.000 | 0.057 | 0.144 | 0.947 | 94.7% |
| Elliptical | 1.000 | 0.588 | 0.567 | 0.547 | 54.7% |

## CONCLUSIONS

SDUV-FEL is well suited for FEL polarization control experiments on the basis of the crossed planar undulators approach. The physical designs indicate that the optimal configuration is that the two crossed undulator keeping as close as possible, and then the final polarization degree could be larger than 90%. By tuning the phase-shifter, fast helicity switching could be also achievable. Up to now, the phase-shifter, the undulator for vertical polarized FEL light, the vacuum chamber modification, and the optical diagnostic sites are under their way. And the experiments are expected to be accomplished at the end of this year.

## ACKNOWLEDGMENT

The authors would like to thank Qika Jia, Yuhui Li, Bart Faatz and Zhirong Huang for useful discussions.